\documentclass[10pt,a4paper]{article}
\usepackage[intlimits]{amsmath}
\usepackage[psamsfonts]{amssymb}
\usepackage{layout}
\usepackage{enumerate}
\usepackage{longtable}
\usepackage{array}

\allowdisplaybreaks
\allowdisplaybreaks 

\renewcommand{\refname}{}

 \allowdisplaybreaks

\begin{document}

\vspace{15mm}


\setcounter{equation}{0} \renewcommand{\refname}{REFERENCES}

\begin{center}
{\Large \textbf{Entangled Optical Solitons in Nonlinear Kerr Dielectric }}\\[%
0pt]
\vspace{1cm}

\textit{Yu.P. Rybakov$^a$ and T.F. Kamalov$^b$}\\[0pt]

\vspace{0.5cm} $^a$ Department of Theoretical Physics\\[0pt]
Peoples' Friendship University of Russia\\[0pt]
117198 Moscow, 6, Mikluho--Maklay str., Russia\\[0pt]
E-mail: soliton4@mail.ru\\[0pt]

$^b$ Physics Department\\[0pt]
Moscow State Opened University\\[0pt]
107996 Moscow, 22, P. Korchagin str., Russia\\[0pt]
E-mail: ykamalov@rambler.ru\\[0pt]
\end{center}

\begin{abstract}
{We consider optical 1D envelope solitons in Kerr dielectric with
cubic nonlinearity and use two--solitons configurations for
modelling entangled states of photons. We calculate spin, momentum
and energy of solitons on the basis of approximate solutions to
the nonlinear Maxwell equations and construct entangled
two--solitons singlet states in special stochastic representation.
}
\end{abstract}

\vspace{1cm}

\section{\textbf{Introduction}}

The interest to the optical envelope solitons in cubic media is widely
known~[1--5]. The main goal of this paper is to show that in the special
``soliton" representation of quantum mechanics~\cite{6} one can use these
soliton solutions to the nonlinear Maxwell equations for modelling entangled
states of photons. The role of the one--particle wave function in this
solitonian scheme is played by the linear combination of solitons with
different phases, the latter being the vector of the special random Hilbert
space advocated by Wiener~\cite{7}. This stochastic representation of the
wave function permits one to construct the entangled states of solitons
modelling the entangled photon states.

\section{\textbf{Main Equations and Structure of Solutions}}

Consider the Kerr nonlinear dielectric, with the permeability $\epsilon $
being the quadratic function of the electric strength $\mathbf{E}$:
\begin{equation}  \label{eq:1}
\epsilon = \epsilon_0 + \epsilon_1 |\mathbf{E}|^2,
\end{equation}
where $\epsilon_0 $ and $\epsilon_1 $ stand for some positive constants. The
corresponding Maxwell equations read:
\begin{equation}  \label{eq:2}
\mbox{rot} \mathbf{E} = - \partial_t \mathbf{B}, \quad \mbox{div} (\epsilon
\mathbf{E}) = 0,
\end{equation}
\vspace{1cm}
\begin{equation}  \label{eq:3}
\mbox{rot} \mathbf{B} = \partial_t \left[\epsilon \mathbf{E}\right], \quad %
\mbox{div} \mathbf{B} = 0,
\end{equation}
where the vacuum velocity of light is chosen as unity: $c = 1$. From (\ref%
{eq:2}) and (\ref{eq:3}) one immediately derives the nonlinear wave equation
for $\mathbf{E}$:
\begin{equation}  \label{eq:4}
\mbox{rot}^2 \mathbf{E} = - \partial^2_t \left[\epsilon(\mathbf{E})\mathbf{E}%
\right].
\end{equation}
Substituting into (\ref{eq:4}) the following vector field:
\begin{equation}  \label{eq:5}
\mathbf{E}_R = \mathbf{e}_R A\,\mbox{sech}(k\xi ), \quad \xi = z - Vt,
\end{equation}
where $\mathbf{e}_R$ stands for the unit vector corresponding to the right
circular polarization:
\begin{equation}  \label{eq:6}
\mathbf{e}_R = \mathbf{e}_x \cos\phi \,+ \mathbf{e}_y \sin\phi, \quad \phi =
\omega t - k_0z,
\end{equation}
one gets three algebraic equations for the constant parameters $A$, $V$, $k$%
, $k_0$, $\omega $:
\begin{equation}  \label{eq:7}
\epsilon_0\left(\omega^2 - k^2V^2\right) = k^2_0 - k^2,
\end{equation}
\vspace{3mm}
\begin{equation}  \label{eq:8}
\epsilon_1A^2\left(9k^2V^2 - \omega^2\right) = 2k^2\left(\epsilon_0 V^2 -
1\right),
\end{equation}
\vspace{3mm}
\begin{equation}  \label{eq:9}
k_0 = \omega V \left(\epsilon_0 + 3\epsilon_1 A^2\right),
\end{equation}
where the natural supposition concerning the envelope pulse was made:
\begin{equation}  \label{eq:10}
k_0 \gg k.
\end{equation}
Introducing the independent parameters $X = k^2/{\epsilon_0 \omega^2}$, $Z =
3A^2\epsilon_1/{\epsilon_0}$, one easily finds from (\ref{eq:7}), (\ref{eq:8}%
) and (\ref{eq:9}):
\begin{gather*}
Z = 3X - 1 + \sqrt{18X^2 + 14X}, \\
\epsilon_0 V^2 = \frac{X + 1}{X + (1 + Z)^2}, \\
k_0^2 = \epsilon_0 \omega^2 (1 + Z)^2\frac{X + 1}{X + (1 + Z)^2}.
\end{gather*}
From the inequality $A > 0$ it is not difficult to estimate the minimal
value of $X$:
\begin{equation*}
X_{\text{min}} \equiv X_0 = \frac{1}{9}\left(- 10 + \sqrt{109}\right)
\approx 0.049.
\end{equation*}
This fact permits to estimate the following useful parameters:
\begin{equation*}
\lambda^2 \equiv \frac{k^2 V^2}{\omega^2} \in \left[\frac{1}{27}, X_0\right]%
; \quad \frac{k^2}{k_0^2} \in \left[X_0, 1\right].
\end{equation*}

Now we intend to search for the magnetic field $\mathbf{B} = \mbox{rot}
\mathbf{A}$. To this end it is necessary to find the transversal vector
potential:
\begin{equation}  \label{eq:11}
\mathbf{A} = - \int\limits^{t}\,\mathbf{E}\,dt.
\end{equation}
The integral in (\ref{eq:11}) can be found via the integration by parts that
gives the asymptotic series as follows:
\begin{equation*}
\omega \int\limits^{t}\,dt\,\cos \phi \,\mbox{sech} (k\xi) = \mbox{sech}
(k\xi)\left[\sin \phi + \lambda \cos \phi\, \mbox{th} (k\xi) + \mathcal{O}%
(\lambda^2)\right].
\end{equation*}
Thus, inserting (\ref{eq:5}) into (\ref{eq:11}) one gets
\begin{equation}  \label{eq:12}
\mathbf{A}_R = \frac{A}{\omega}\mbox{sech} (k\xi)\left[\mathbf{e}_L -
\mathbf{e}_R \lambda \,\mbox{th}(k\xi ) + \mathcal{O}(\lambda^2)\right].
\end{equation}
From (\ref{eq:12}) one easily finds the magnetic field
\begin{equation}  \label{eq:13}
\mathbf{B}_R = \frac{A}{\omega}\mbox{sech} (k\xi)\left[\mathbf{e}_L\left(k_0
- k\lambda + 2k\lambda \,\mbox{th}^2(k\xi)\right) + \mathbf{e}_R (k -
\lambda k_0)\,\mbox{th}(k\xi ) + \mathcal{O}(\lambda^2)\right].
\end{equation}
It is worth-while to underline that the soliton solution corresponding to
the left circular polarization can be obtained from (\ref{eq:5}) and (\ref%
{eq:12}) via the transposition
\begin{equation*}
\mathbf{e}_R \Longrightarrow \mathbf{e}^{\prime}_L; \quad \mathbf{e}_L
\Longrightarrow \mathbf{e}^{\prime}_R,
\end{equation*}
where the denotation is used:
\begin{equation*}
\mathbf{e}^{\prime}_R = \mathbf{e}_x \cos\phi \,- \mathbf{e}_y \sin\phi
\quad \mathbf{e}^{\prime}_L = \mathbf{e}_x \sin\phi \,+ \mathbf{e}_y
\cos\phi .
\end{equation*}

\section{\textbf{Physical Observables and Entangled States}}

Using the solution found it is possible to calculate the integrals of motion
describing the soliton configuration, that is the physical observables: the
energy $W$, the spin $\mathbf{S}$ and the momentum $\mathbf{P}$. These
observables can be constructed via the Lagrangian density
\begin{equation*}
\mathcal{L}=\frac{1}{4}\left( 2\epsilon _{0}\mathbf{E}^{2}+\epsilon _{1}%
\mathbf{E}^{4}-2\mathbf{B}^{2}\right)
\end{equation*}%
if one follows the standard variational procedure:
\begin{equation}
W=\frac{1}{4}\int \,dz\,\left( 2\epsilon _{0}\mathbf{E}^{2}+3\epsilon _{1}%
\mathbf{E}^{4}+2\mathbf{B}^{2}\right) ,  \label{eq:14}
\end{equation}%
\vspace{3mm}
\begin{equation}
\mathbf{S}=\int \,dz\,\epsilon \lbrack \mathbf{E}\mathbf{A}],  \label{eq:15}
\end{equation}%
\vspace{3mm}
\begin{equation}
P_{z}=\int \,dz\,\epsilon \left( \mathbf{E}\partial _{z}\mathbf{A}\right) .
\label{eq:16}
\end{equation}%
Inserting (\ref{eq:5}) and (\ref{eq:13}) into (\ref{eq:14}), (\ref{eq:15})
and (\ref{eq:16}) one gets
\begin{equation}
W\approx \frac{A^{2}}{k}\left[ \epsilon _{0}+\epsilon _{1}A^{2}+\frac{1}{%
3\omega ^{2}}\left( 3k_{0}^{2}-4k\lambda +k^{2}\right) \right] ,
\label{eq:17}
\end{equation}%
\vspace{3mm}
\begin{equation}
\mathbf{S}_{R}=-\mathbf{S}_{L}=\mathbf{e}_{z}S,\quad S\approx \frac{2A^{2}}{%
3k\omega }\left( 3\epsilon _{0}+2\epsilon _{1}A^{2}\right) ,  \label{eq:18}
\end{equation}%
\vspace{3mm}
\begin{equation}
\mathbf{P}=\mathbf{e}_{z}P,\quad P=k_{0}S.  \label{eq:19}
\end{equation}

Now we define the subsidiary complex vector function
\begin{equation}  \label{eq:20}
\mathbf{\varphi} = \frac{1}{\sqrt{2}}\left(\nu \mathbf{A} + \frac{\imath}{\nu%
}\mathbf{\pi}\right), \quad \mathbf{\pi} = - \epsilon \mathbf{E},
\end{equation}
where the constant $\nu$ is defined by the normalization condition
\begin{equation}  \label{eq:21}
\hbar = \int\,dz\, |\mathbf{\varphi}|^2,
\end{equation}
$\hbar$ being the Planck constant. Stochastic representation of the
one-particle wave function $\Psi_N$ can be defined as the linear combination
of the functions (\ref{eq:20}) determined in $N$ independent trials:
\begin{equation}  \label{eq:22}
\Psi_N (t,z) = (\hbar N)^{-1/2} \sum_{j = 1}^{N} \mathbf{\varphi}_j (t, z).
\end{equation}
The index $j$ in (\ref{eq:22}) runs over the set of independent (random)
trials, the number $N$ of which is supposed to be very large (the frequency
hypothesis by von Mises). Thus, the formula (\ref{eq:22}) gives the
stochastic realization of the wave function via the infinite dimensional
objects---solitons.

To show that $\Psi_N $ plays the role of the probability amplitude, it is
sufficient to calculate the integral
\begin{equation}  \label{eq:23}
\rho_N = \frac{1}{\triangle \vee }\, \int\limits_{\triangle \vee \subset
\mathbb{R}^1}dz \,\left|\Psi_N\right|^2,
\end{equation}
which is taken over the small volume (interval) $\triangle \vee \gg \vee_0 $%
, where $\vee_0 \sim 1/k$ stands for the proper volume (size) of the
soliton. It can be easily shown~\cite{6} that with the probability $P = 1 -
\alpha \vee_0/{\triangle \vee } $, $\alpha \sim 1 $, the integral (\ref%
{eq:23}) equals to
\begin{equation*}
\rho_N = \frac{\triangle N}{N\triangle \vee },
\end{equation*}
where $\triangle N $ is the number of trials, for which the centers of
particles---solitons appear to be contained in $\triangle \vee $. That is
why $\rho_N $ can be identified with the coordinate probability density.

Now let us consider the measuring procedure for some observable $A$
corresponding, due to E.~Noether's theorem, to the symmetry group generator $%
\widehat{M}_{A}$. For example, the momentum $\mathbf{P}$ is related with the
generator of space translation $\widehat{M}_{P}=-\imath \,\bigtriangledown $%
, the angular momentum $\mathbf{L}$ is related with the generator of space
rotation $\widehat{M}_{L}=\mathbf{J}$ and so on. As a result one can
represent the classical observable $A_{j}$, first for the $j$--th trial
performed in the experiment with $3D$ particle--soliton, in the form
\begin{equation*}
A_{j}=\int \,d^{3}x\,\pi _{j}\imath \widehat{M}_{A}\phi _{j}=\int
\,d^{3}x\,\varphi _{j}^{\ast }V\varphi _{j},
\end{equation*}%
with $\pi _{j}$ being the generalized momentum in $j$--th trial and $\phi
_{j}$ being the generalized coordinate (field). In our case $\phi _{j}$
corresponds to $\mathbf{A}_{j}$ and $\pi _{j}$ corresponds to $\mathbf{\pi }%
_{j}=-\epsilon \mathbf{E}_{j}$. The corresponding mean value reads
\begin{eqnarray}
\mathbb{E}(A) &\equiv &\frac{1}{N}\sum_{j=1}^{N}A_{j}=\frac{1}{N}%
\sum_{j=1}^{N}\int \,d^{3}x\,\varphi _{j}^{\ast }\widehat{M}_{A}\varphi _{j}
\notag \\
&=&\int \,d^{3}x\,\Psi _{N}^{\ast }\widehat{A}\Psi _{N}+O(\frac{\vee _{0}}{%
\triangle \vee }),  \label{eq:24}
\end{eqnarray}%
where the Hermitian operator $\widehat{A}$ stands for $\widehat{A}=\hbar
\widehat{M}_{A}$. Thus, up to the terms of the order $\vee _{0}/\triangle
\vee \ll 1$ we obtain the standard quantum mechanical rule for calculating
the mean values of observables~[8], [9]. Similar expressions can be found
for our $1D$ particles--solitons. For example, the spin operator has the
standard form
\begin{equation*}
\left( \widehat{S}_{k}\right) _{lm}=-\imath \hbar \epsilon _{klm},
\end{equation*}%
i. e. the rotation matrix in $\mathbb{R}^{3}$.

Now we consider two-solitons singlet states, that is construct the entangled
solitons configuration with the zero spin and momentum:
\begin{equation}  \label{eq:25}
\varphi ^{(12)}(t, z_1, z_2) = \frac{1}{\sqrt{2}}\left[\mathbf{\varphi}_L
(t, - z_1) \otimes \mathbf{\varphi}_R(t, z_2) - \mathbf{\varphi}_R (t, -
z_1) \otimes \mathbf{\varphi}_L(t, z_2)\right].
\end{equation}
The normalization condition for the configuration (\ref{eq:25}) reads
\begin{equation*}
\int\,dz_1\,\int\,dz_2\,\varphi ^{(12)*}\varphi ^{(12)} = {\hbar }^2.
\end{equation*}
Now it is not difficult to find the expression for the stochastic wave
function for the singlet two--solitons entangled state:
\begin{equation}  \label{eq:26}
\Psi _N \left(t, z_1, z_2\right) = {\left({\hbar }^{2}N\right)}%
^{-1/2}\sum\limits_{j=1}^{N}\,\varphi_j^{(12)},
\end{equation}
where $\varphi_j^{(12)}$ corresponds to the entangled solitons configuration
in the $j$--th trial.

In conclusion we express the hope that this formalism of entangled solitons
can be suitable for modelling the real photons.

\end{document}